\documentclass[runningheads]{llncs}
\usepackage{subfigure}
\usepackage[frozencache=true,cachedir=_minted-main]{minted}
\usepackage{multirow}
\usepackage{tabularx}
\usepackage{wrapfig}
\usepackage{xspace}
\usepackage{verbatim}
\usepackage{graphicx}

\newcommand{\system}{\texttt{AutoMPHC}\xspace}
\newcommand{\lang}{Intrepydd\xspace}

\newcommand{\tabspace}[1]{~{#1}~}
\newcommand{\tts}[1]{\texttt{\small #1}}
\newcommand{\ttt}[1]{\texttt{#1}}

\titlerunning{\system{}}
\authorrunning{Shirako, Hayashi, Paul, Tumanov, Sarkar}

\begin{document}
\title{Automatic Parallelization of Python Programs for Distributed Heterogeneous Computing}

\author{
  Jun Shirako
  \and
  Akihiro Hayashi
  \and
  Sri Raj Paul
  \and
  \\ Alexey Tumanov
  \and
  Vivek Sarkar
}

\institute{School of Computer Science, Georgia Institute of Technology}

\maketitle

\begin{abstract}
This paper introduces a novel approach to automatic ahead-of-time (AOT) parallelization and optimization of sequential Python programs for execution on distributed heterogeneous platforms.  Our approach enables AOT source-to-source transformation of Python programs, driven by the inclusion of type hints for function parameters and return values.  These hints can be supplied by the programmer or obtained by dynamic profiler tools; multi-version code generation guarantees the correctness of our AOT transformation in all cases.

Our compilation framework performs automatic parallelization and sophisticated high-level code optimizations for the target distributed heterogeneous hardware platform.  It includes extensions to the polyhedral framework that unify user-written loops and implicit loops present in matrix/tensor operators, as well as automated section of CPU vs. GPU code variants.  Further, our polyhedral optimizations enable both intra-node and inter-node parallelism.  Finally, the optimized output code is deployed using the Ray runtime for scheduling distributed tasks across multiple heterogeneous nodes in a cluster.

Our empirical evaluation shows significant performance improvements relative to sequential Python in both single-node and multi-node experiments, with a performance improvement of over 20,000$\times$ when using 24 nodes and 144 GPUs in the OLCF Summit supercomputer for the Space-Time Adaptive Processing (STAP) radar application.

\end{abstract}

\section{Introduction}
Multiple simultaneous disruptions are currently under way in both
hardware and software, as we consider the implications for future
parallel systems.  In hardware, “extreme heterogeneity” has become
critical to sustaining cost and performance improvements with the end
of Moore’s Law, but poses significant productivity challenges for
developers.  In software, the rise of large-scale data science and AI
applications is being driven by domain scientists from diverse
backgrounds who demand the programmability that they have come to
expect from high-level languages like Python~\cite{python}.  While
this paper focuses on Python as an exemplar of modern high-productivity
programming, the approach in this paper is equally applicable to other
high-productivity languages such as Julia~\cite{julia}.

A key challenge facing domain scientists is determining how to enable their
Python-based applications to use the parallelism inherent in both distributed
and heterogeneous computing.  A typical workflow for domain
scientists is to experiment with new algorithms by starting with smaller
datasets and then moving on to larger datasets.  A tipping point is
reached when the dataset size becomes too large to be processed within
a single node, and another tipping point is reached when
there is a need to use accelerators such as GPUs.

One approach to dealing with these tipping points is to rely on
experienced programmers with a deep (“ninja level”) expertise in
computer architecture and code optimization for accelerators and
inter-node communication. However, this approach is a non-starter for
many domain scientists due to its complexity and the skills required.  For example, even though
Python bindings for MPI~\cite{mpi4py} have been available for many
years, there has been very little adoption of these bindings by domain
scientists.  An alternate approach is to augment a high-productivity
language with native libraries that include high-performance
implementations of commonly used functions, e.g., functions in the
NumPy~\cite{numpy} and SciPy~\cite{scipy} libraries for Python. However, fixed
library interfaces and implementations do not address the needs
of new applications and algorithms.  Yet another approach is to develop and use
Domain Specific Languages (DSLs); this approach has recently begun
showing promise for certain target domains, e.g., PyTorch and
TensorFlow for machine learning, Halide for image processing
computations, and TACO for tensor kernels. However, the
deliberate lack of generality in DSLs poses significant challenges
in requiring
domain
scientists to learn multiple DSLs and to integrate DSL kernels into their overall programming
workflow, while also addressing corner cases that may not be supported
by DSLs.

In this paper, we make the case for new
advances to enable productivity and programmability of future HPC
platforms for domain scientists.
The goal of our system, named \system{},
is Automation of Massively Parallel and Heterogeneous Computing,
obtained 
by delivering the benefits of distributed heterogeneous hardware
platforms to domain scientists without requiring them to undergo any
new training.  As a first step towards this goal, this paper 
introduces a novel approach to automatic ahead-of-time (AOT)
parallelization and optimization of sequential Python programs for
execution on distributed heterogeneous platforms, and supports program multi-versioning
for specializing code generation to different input data types and
different target processors.  The optimized code is deployed using the
Python-based Ray runtime~\cite{Ray18} for scheduling distributed tasks across multiple heterogeneous nodes in a cluster.

As a simple illustration of our approach, consider two versions of the
PolyBench~\cite{10.1145/3446804.3446842} {\tt correlation} benchmark shown in 
Figures~\ref {fig:corr-list} and \ref{fig:corr-numpy}.  The first case
represents a list-based pattern implemented using three
explicit Python loops that access elements of lists (as surrogates for
arrays), which might have been written by a domain scientist familiar
with classical books on algorithms such as \cite{numerical-recipes}.
The second case represents
a NumPy-based pattern with one explicit loop and a two-dimensional array statement in line~7 of
Figure~\ref{fig:corr-numpy}, which might have been written by a domain
scientist familiar with matrix operations.
A unique feature of our approach is the ability to support both
explicit Python
loops and implicit loops from NumPy operators and library calls in a unified
optimization framework.
Table~\ref{tbl:corr-large-timing} shows that the NumPy-based version
of the  {\tt correlation} benchmark
performs better than the list version, while our
approach (which can be applied to either style of input) performs significantly better than both.  Additional
performance results are discussed in Section~\ref{sec:experiment}.

\begin{figure}[t]
\begin{minted}[linenos,fontsize=\scriptsize]{python}
def kernel(self, float_n: float, data: list, corr: list, mean: list, stddev: list):
    ...
    for i in range(0, self.M-1):
        corr[i][i] = 1.0
        for j in range(i+1, self.M):
            corr[i][j] = 0.0
            for k in range(0, self.N):
                corr[i][j] += (data[k][i] * data[k][j])
            corr[j][i] = corr[i][j]
    corr[self.M-1][self.M-1] = 1.0
\end{minted}
\vspace{-5mm}
\caption{PolyBench-Python correlation: List version (default)}
\label{fig:corr-list}
\begin{minted}[linenos,fontsize=\scriptsize]{python}
from numpy.core.multiarray import ndarray
...
def kernel(self, float_n: float, data: ndarray, corr: ndarray, mean: ndarray, stddev: ndarray):
    ...
    corr[np.diag_indices(corr.shape[0])] = 1.0
    for i in range(0, self.M - 1):
        corr[i,i+1:self.M] = (data[0:self.N,i] * data[0:self.N,i+1:self.M].T).sum(axis=1)
    tril_indices = np.tril_indices( n=self.M, m=self.M, k=-1 )
    corr[tril_indices] = corr[triu_indices]
    corr[self.M - 1, self.M - 1] = 1.0
\end{minted}
\vspace{-5mm}
\caption{PolyBench-Python correlation: NumPy version}
\label{fig:corr-numpy}
\end{figure}

\begin{table}[t]
\begin{center}
\caption{Execution time of correlation (dataset = large)}
\label{tbl:corr-large-timing}
{\small
\begin{tabular}{|c|c|c|}
\hline
\tabspace{List version} & \tabspace{NumPy version} & \tabspace{Our optimization (Figure~\ref{fig:corr-poly}c)} \\
\hline
152.5 [sec] & 2.212 [sec] & 0.07163 [sec] \\
\hline
\end{tabular}
}
\end{center}
\end{table}

In summary, this paper makes the following contributions:
\begin{itemize}
\item A novel approach to automatic ahead-of-time (AOT)
  parallelization and optimization of sequential Python programs for
  execution on distributed heterogeneous platforms.  Our approach is
  driven by the inclusion of type hints for function parameters and
  return values, which can be supplied by the programmer or obtained by dynamic profiler tools; multi-version code generation guarantees the correctness of our AOT transformation in all cases.
\item Automatic parallelization and high-level code optimizations for
  the target distributed heterogeneous hardware platform, based on
  extensions to the polyhedral framework that unify user-written loops
  and implicit loops present in matrix/tensor operators, as well as
  automated selection of CPU vs. GPU code variants.
 \item Automatic code generation for targeting the Ray runtime to schedule distributed tasks across multiple heterogeneous nodes in a cluster.
 \item An empirical evaluation for 15 Python-based benchmarks from the Polybench
   suite on a single node with multiple GPUs, and another evaluation
   of the Space-Time Adaptive Processing (STAP) radar
   application in Python.  Both evaluations  show significant performance
  improvements due to the use of \system{}.  In the case of STAP, the performance improvement 
  relative to the original Python code was over 20,000$\times$ when using 24
  nodes and 144 GPUs (6 GPUs/node) in the OLCF Summit supercomputer.
\end{itemize}
\label{sec:introduction}

\section{Background}
\subsection{Intrepydd Compiler}
The \lang programming language~\cite{Intrepydd-onward20} introduced a
subset of Python that is amenable to ahead-of-time (AOT) compilation into C++.  It is 
intended for writing kernel functions rather than complete or main programs.  The
C++ code generated from \lang kernels can be imported into a Python
application or a C++ application.

A key constraint in the \lang subset of Python is the requirement
that \lang function definitions include type annotations for parameters and
return values.  Given these type annotations, the
\lang compiler statically infers the types of local variables and
expressions.  The \lang{} tool chain includes a library knowledge base, which specifies type rules for a wide range of standard library
functions used by Python programs. 
As discussed in the following sections, the \system{} system extends the \lang{} tool chain to serve as a Python-to-Python optimization and parallelization system; there is no C++ code generated by \system{}.

It is important to note that \lang{} also includes  extensions to standard Python to enable C++ code generation.  These extensions include statements with explicit parallelism (e.g., {\tt pfor} for parallel loops) and special library functions.  In contrast, \system{} does not rely on any of these extensions.  All input code to \system{} and all output code generated by \system{} can be executed on standard Python implementations.

\subsection{\system{}  Runtime}
\begin{figure}
  \begin{center}
    \includegraphics[width=0.7\textwidth]{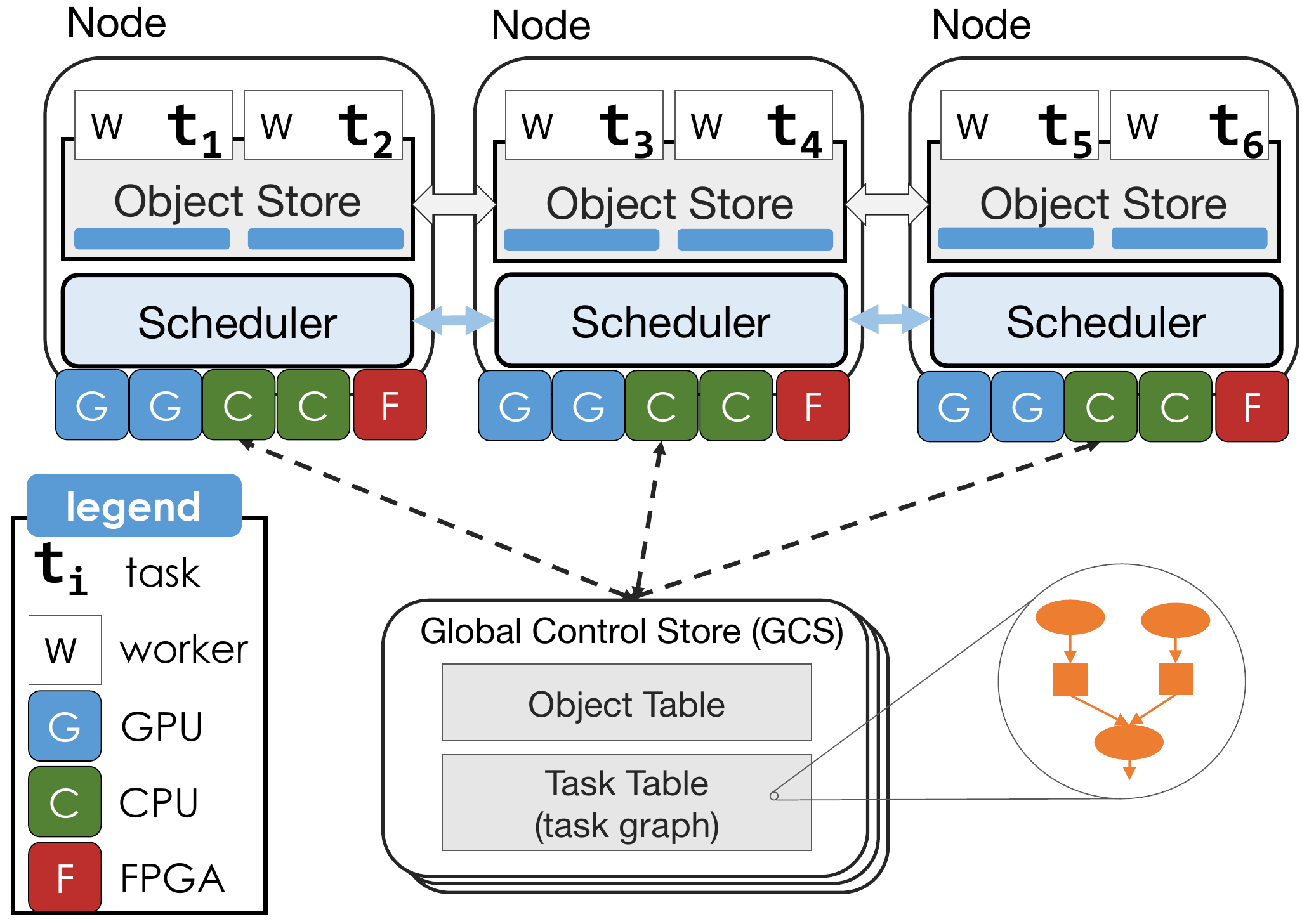}
  \end{center}
  \vspace{-5mm}
  \caption{\system{} distributed runtime architecture.}
  \label{fig:ray-arch}
\end{figure}
We use Ray~\cite{Ray18} as the base distributed runtime framework. Ray features a number of properties beneficial for \system{}.
First, the ability to simultaneously support both {\em stateless} and {\em stateful} computation---one of its key research contributions useful for a heterogeneous mix of CPU and GPU compute.
Stateless computation, in the form of side effect free tasks, is best suited for processing large data objects or partitions on numerous CPU resources. Stateful computation is beneficial for GPU tasks.
We create tasks for this distributed runtime by \textit{automatically} compiling chunks of code into Ray tasks.
Each Ray task then can be spawned asynchronously.
A full directed acyclic graph (DAG) of such task instantiations is dynamically constructed and submitted for execution without waiting for intermediate computation results. It enables \system{} to 
(a) hide the latency of task instantiation and propagation to workers for execution, 
(b) extract pipeline parallelism, and,
(c) extract parallelism from the partial order of the dynamically constructed directed acyclic task graph.
As Ray tasks are instantiated,
they return immediately with a future-like construct, called an ObjectID --- an object handle that refers to a globally addressable object.
The object is eventually fulfilled and can be extracted with a blocking \tts{ray.get(object\_id)} API. We note that the distributed object store (Fig.~\ref{fig:ray-arch} used for the lifecycle of these objects is immutable---a property that elides the need for expensive consistency protocols, state coherence protocols, and other synchronization overheads needed for data correctness. Critically, this alleviates the need for expensive MPI-style distributed barriers and, therefore, does not suffer from the otherwise common straggler challenges---an important property for heterogeneous compute at scale.
Finally, data store immutability, combined with the deterministic nature of the task graph, enable fault tolerance, as any missing object in the graph can be recomputed by simply replaying the sub-graph leading up to and including the object's parent vertex. This mechanism can be triggered automatically and comes with minimal overhead on the critical path of a task~\cite{lineage-sosp19}.

\section{Overview of our Approach}
\begin{figure}[t]
\begin{center}
\includegraphics[width=0.8\columnwidth]{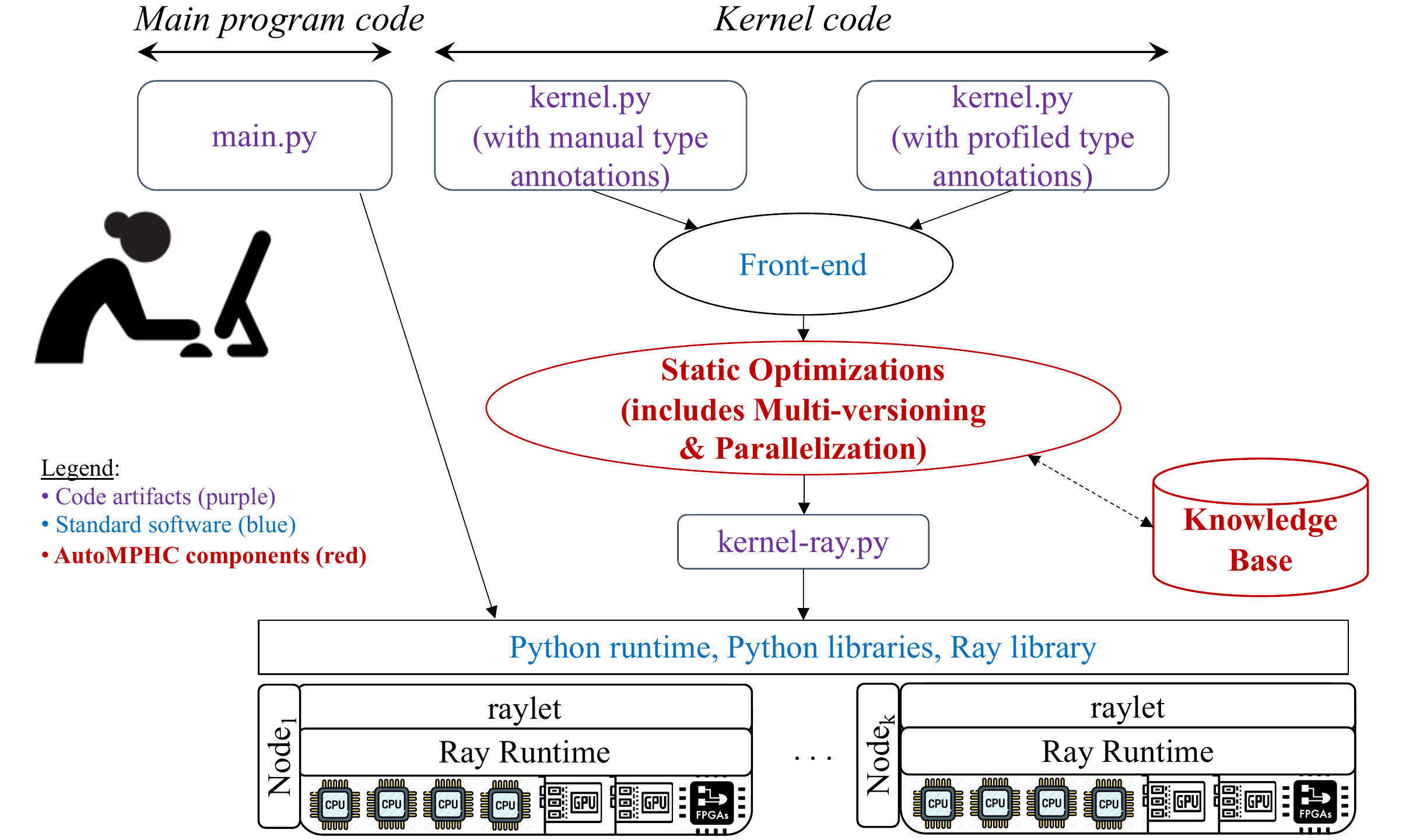}
\end{center}
\vspace{-5mm}
\caption{Overall Design of \system system}
\label{fig:overall}
\end{figure}

Figure~\ref{fig:overall} summarizes the overall design of our proposed \system{} system.  User-developed code  is a combination of {\em main program code} and {\em kernel code}, where the former is executed unchanged and the latter is optimized by \system{} via automatic ahead-of-time (AOT) source-to-source transformations.  There are two forms of kernel code supported by our system --- one in which type annotations are manually provided by the user, and another in which type annotations are obtained by a type profiler such as MonkeyType.  In both cases, the type annotations serve as {\em hints} since the multi-version code generation guarantees the correctness of our AOT transformations in all cases (whether or not the actual inputs match the type annotations).

The kernel functions with type annotations (hints) are first translated by the Front-end to an Abstract Syntax Tree (AST) representation implemented using the standard Python Typed AST package.  The core optimizations in \system{} are then performed on the AST, including multi-version code specialization (Section~\ref{sec:multi-versioning}), polyhedral optimizations (Section~\ref{sec:polyhedral}), and generation of distributed parallel code using Ray tasking APIs along with generation of heterogeneous code using selective NumPy-to-CuPy conversion (Section~\ref{sec:numpy-to-cupy}).  These Static Optimizations benefit from the use of the \system{} Knowledge Base, which includes dataflow and type information for many commonly used library functions.   The transformed code is then executed on a distributed heterogeneous platform using standard Python libraries in addition to Ray.

\section{Optimizations}
The \system compiler is the extension of \lang
compiler~\cite{Intrepydd-onward20}, which supports type inference and
basic optimizations including loop invariant code motion, sparsity
optimization, and array allocation/slicing optimizations.  In the
following sections, we present newly developed optimizations for
automatic parallelization targeting distributed heterogeneous systems.

\subsection{Program Multi-versioning for Specialized Code Optimizations}\label{sec:multi-versioning}
Multi-versioning is an approach to data-aware optimizations, which
generates multiple code versions specialized under certain conditions
at compile-time and selects a proper code version at runtime.  In our
framework, we consider two classes of conditions, {\em legality-based}
and {\em profitability-based}.  All the conditions are organized as
decision trees, where legality conditions are located at higher levels
while profitability conditions are at lower levels in general.

The legality conditions are mainly used to verify the data type
annotations attached on function parameters and returns.  In our
approach, the type annotations are used as hints and can be different
from actual types given at runtime.  Further, the correctness of array
rank/dimensionality inference is critical to the polyhedral
optimizations (Section~\ref{sec:polyhedral}).  The multi-versioning
serves as runtime checks of annotated/inferred types and ranks for
specialized code version while ensuring correct behavior for others,
as shown in Figure~\ref{fig:multiver}.

\begin{figure}[t]
\begin{minted}[linenos,fontsize=\scriptsize]{python}
def kernel(self, float_n: float, data: ndarray, ...):
    if type(float_n) == float and type(data) == ndarray and ...:
        if data.ndim == 2 and ... :
            ...  # Code with type-specific and rank-specific optimizations
        else:
            ...  # Code with type-specific optimizations
    else:
        ...  # Code without type-specific optimizations
\end{minted}
\vspace{-5mm}
\caption{Multi-versioning for PolyBench-Python correlation}
\label{fig:multiver}
\end{figure}

The profitability conditions can cover a broad range of
conditions/scenarios related to runtime performance rather than
correctness.  As described later, the \system compiler can generate
two versions of optimized kernels, one for CPUs and the other for
GPUs.  The runtime condition between these two versions is a typical
example of profitability conditions (Section~\ref{sec:numpy-to-cupy}).

\subsection{Polyhedral Optimizations}\label{sec:polyhedral}
Polyhedral compilation has provided significant advances in the
unification of affine loop transformations combined with powerful code
generation
techniques~\cite{Bondhugula:2016:PAP:2914585.2896389,Verdoolaege:2013:PPC:2400682.2400713,Zinenko:2018:MCD:3178372.3179507}.
However, despite these strengths in program transformation, the
polyhedral frameworks lack support for: 1) dynamic control flow and
non-affine access patterns; and 2) library function calls in general.
To address the first limitation, we have extended the polyhedral
representation, Static Control Parts (SCoP), to represent unanalyzable
expressions as a compound ``black-box'' statement with approximated
input/output relations.  To address the second limitation, we took
advantage of our library knowledge base to obtain element-wise
dataflow relations among function arguments and return values, whose
examples are shown in Table~\ref{tbl:library-summary-amphc}.  These
unique features enable the co-optimization of both explicit loops and
implicit loops from operators and library calls in a unified
optimization framework, as detailed in the following sections.

Given SCoP representation extracted from the Python IR, the \system
polyhedral optimizer, which is built on PolyAST~\cite{ShPS14,ShHS17}
framework, computes dependence constraints and performs program
transformations.  Finally, the optimized SCoP representation is
converted back to Python IR with the help of library knowledge base
for efficient library mapping.

\begin{table}[t]
\begin{center}
\caption{NumPy examples in library knowledge base}
\label{tbl:library-summary-amphc}
\begin{tabular}{|l|c|l|}
\hline
\tabspace{\bf Library function} & \tabspace{\bf Domain} & \tabspace{\bf Semantics and dataflow} \\
\hline
\tabspace{transpose$_{2D}$} & \tabspace{$(i_0, i_1)$} & \tabspace{$R[i_0, i_1]$ {\bf := }$A_1[i_1, i_0]$} \\
\hline
\tabspace{mult$_{1D,2D}$} & \tabspace{$(i_0, i_1)$} & \tabspace{$R[i_0, i_1]$ {\bf := }$A_1[i_1] \times A_2[i_0, i_1]$} \\
\hline
\tabspace{sum$_{1D}$} & \tabspace{$(0)$} & \tabspace{$R$ {\bf := }$\Sigma_k A_1[k]$} \\
\hline
\tabspace{sum$_{2D,axis=1}$} & \tabspace{$(i_0)$} & \tabspace{$R[i_0]$ {\bf := }$sum_{1D}${\bf (}$A_1[i_0, :]${\bf )}} \\
\hline
\tabspace{dot$_{2D,2D}$} & \tabspace{$(i_0, i_1)$} & \tabspace{$R[i_0, i_1]$ {\bf := }$sum_{1D}${\bf (}$mult_{1D,1D}${\bf (}$A_1[i_0,:]${\bf , }$A_2[:,i_1]${\bf ))}} \\
\hline
\tabspace{fft$_{2D,axis=1}$} & \tabspace{$(i_0)$} & \tabspace{$R[i_0,:]$ {\bf := }$\mathit{fft_{1D}}${\bf (}$A_1[i_0, :]${\bf )}} \\
\hline
\end{tabular}
\end{center}
\end{table}

\begin{figure}[t]
\begin{center}
\includegraphics[width=\columnwidth]{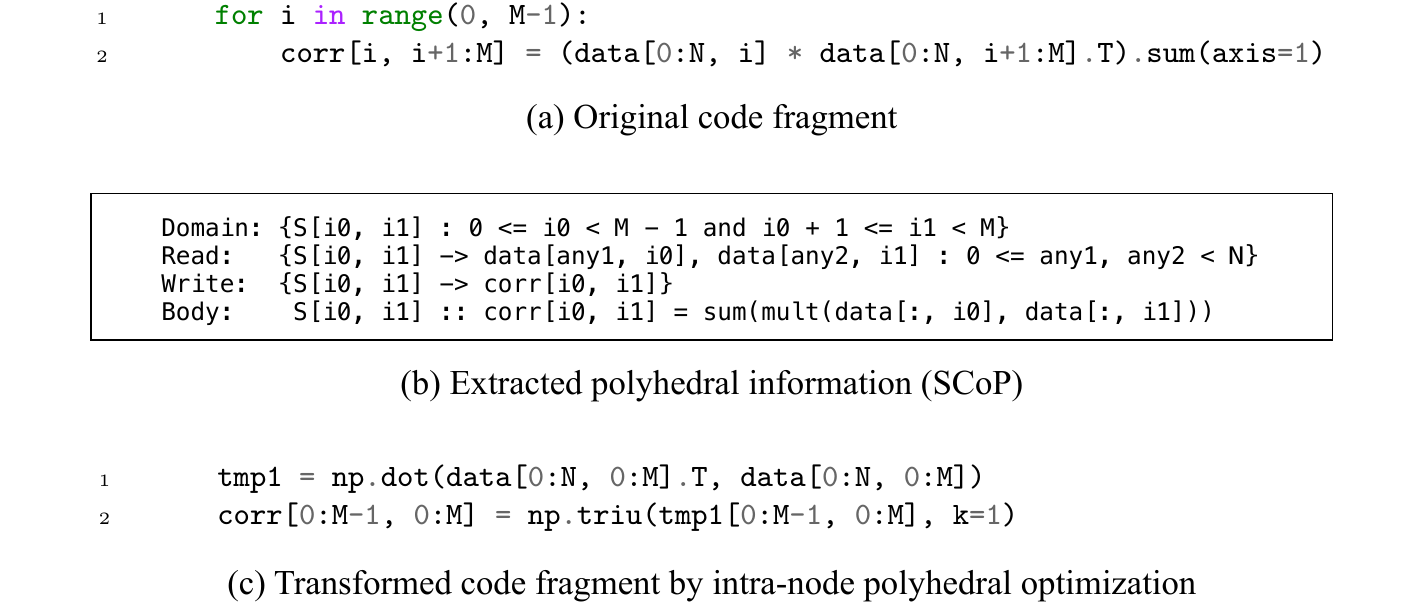}
\end{center}
\vspace{-5mm}
\caption{Kernel from the PolyBench-Python correlation}
\label{fig:corr-poly}
\end{figure}

\noindent
{\bf Intra-node parallelization:}
The optimization policy for intra-node level is to provide sufficient
parallelism to fully utilize the efficient multithreaded
implementations such as BLAS-based NumPy and CuPy.  Our modified
PolyAST~\cite{ShPS14} algorithm applies loop distribution to split
different library calls while maximizing the iteration domain (i.e.,
amount of computation) that can be mapped to a single library function
call.
The SCoP-to-Python-IR generation stage leverages the library knowledge
base to select the efficient combination of available library
functions for each statement whenever possible.  The maximal matching
strategy is currently employed if multiple choices are available.

Figure~\ref{fig:corr-poly}a shows the computationally heavy code
fragment of PolyBench-Python \ttt{correlation} NumPy version, which
has a \ttt{for} loop enclosing a sequence of NumPy function calls: 2-D
array transpose overlapping \ttt{T} operator; 1-D$\times$2-D array
multiply overlapping \ttt{*} operator, and 2-D array
summation \ttt{sum} to produce 1-D result.
Based on the type inference results, the polyhedral phase first
identifies these library functions with specific types and array
ranks.  As summarized in Table~\ref{tbl:library-summary-amphc}, the
library knowledge base provides the element-wise dataflow information
and operational semantics of these functions, which are used to
extract the SCoP information and semantics of each statement
(Figure~\ref{fig:corr-poly}b).  Both explicit and implicit loops are
unified in a triangular iteration domain; and the same loop order is
selected by the transformation stage.
Given statement body of \tts{sum(mult(data[:, i0], data[:, i1]))}, the
SCoP-to-Python-IR generation stage selects the combination of
matrix-matrix multiplication \ttt{numpy.dot} and 2-D transpose \ttt{T}
as the best mapping, followed by \ttt{numpy.triu} to update only the
triangular iteration domain (Figure~\ref{fig:corr-poly}c).
As evaluated in Section~\ref{sec:result-single}, this transformation
sufficiently increases the intra-node parallelism per library call and
resulted in significant improvements for several benchmarks.

When the input program is written only with explicit loops, e.g., List
version in Figure~\ref{fig:corr-numpy}, our approach extracts similar
SCoP and generates the same code with additional conversions between
List and NumPy array.

\begin{figure}[t]
\begin{center}
\includegraphics[width=\columnwidth]{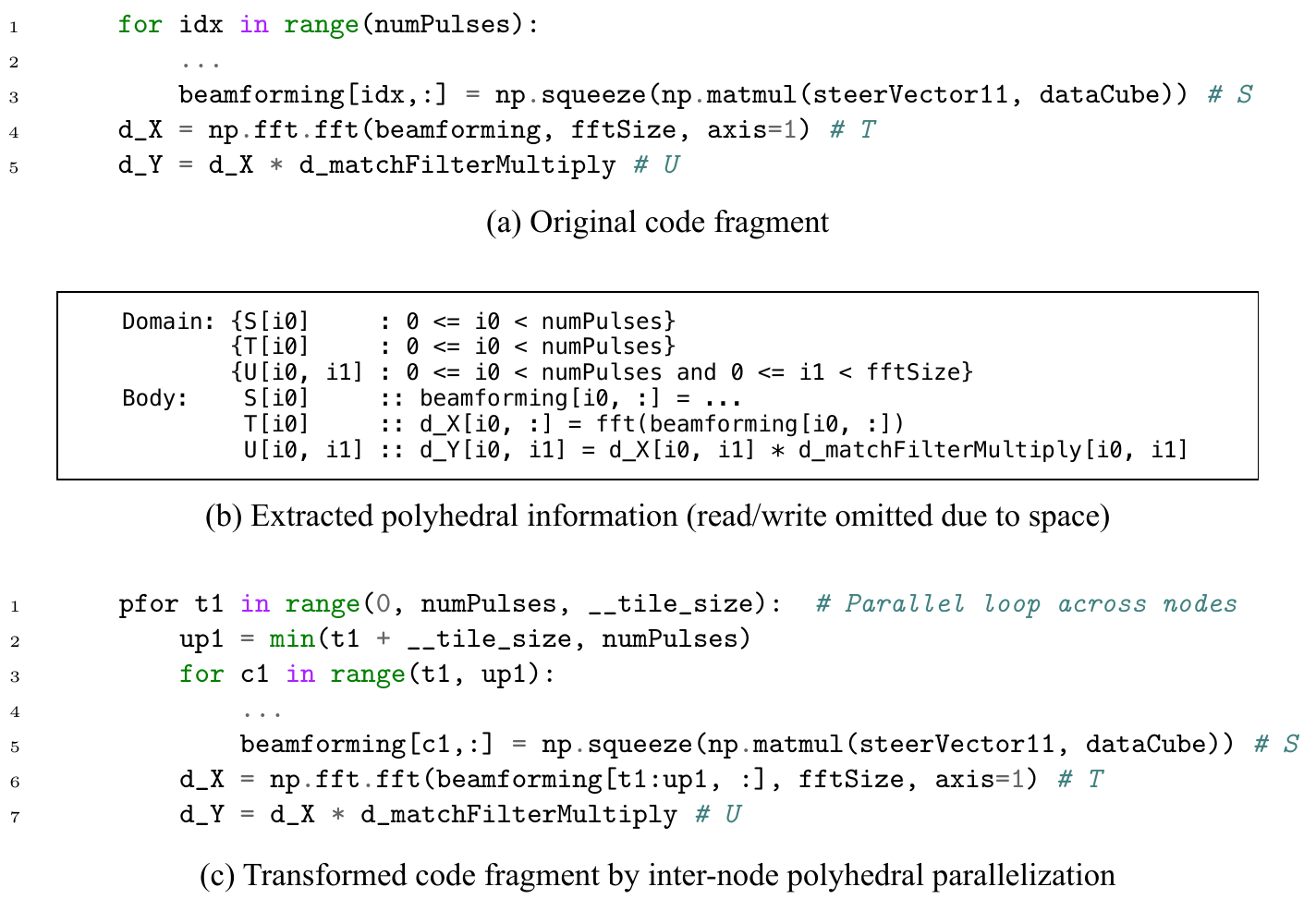}
\end{center}
\vspace{-5mm}
\caption{Kernel from the STAP Signal Processing Application}
\label{fig:stap-poly}
\end{figure}

\noindent
{\bf Inter-node parallelization:}
The optimization policy for inter-node level is equivalent to the
original PolyAST~\cite{ShPS14} algorithm that maximizes outermost
level parallelism, while incorporated with our data layout
transformation approach~\cite{ShSa19,Shirako-lcpc2020} to reduce the
total allocated array sizes and data movement across Ray tasks.
Analogous to the GPU two-level parallelization~\cite{ShHS17}, our
polyhedral optimizer selects different {\em schedules} -- i.e.,
compositions of loop transformations and parallelization -- for
inter-node and intra-node levels individually; and integrates them
into the final schedule via loop tiling.

Figures~\ref{fig:stap-poly}a and \ref{fig:stap-poly}b respectively
show the computational kernel of the STAP radar application and the
extracted SCoP information.  The explicit loop with statement S and
the \ttt{fft} call of statement T are handled as 1-D iteration domains
while 2-D$\times$2-D array multiply of statement U is handled as a 2-D
iteration domain.  The polyhedral optimizer identifies the outermost
level parallelism and computes the inter-node schedule that fuses
these statements into a single parallel loop.  The transformed code
after integrating the inter-node and intra-node schedules is shown in
Figure~\ref{fig:stap-poly}c, where \ttt{pfor} is the parallel loop
construct to be distributed across Ray tasks.

\subsection{NumPy-to-CuPy Conversion and Parallelized Code Generation}\label{sec:numpy-to-cupy}
After the polyhedral phase, the program multi-versioning
(Section~\ref{sec:multi-versioning}) is applied to the {\em pfor}
parallel loops and generates both sequential and parallel versions.
The profitability condition, which makes the decision on whether the
loop to be distributed across nodes via Ray runtime, is generated by a
simple cost-based analysis and summarized as a threshold expression
using loop counts.
This analysis also includes the feasibility and profitability check of
the CuPy conversion for given sequence of NumPy library calls.  The
current implementation takes an all-or-nothing approach for the
NumPy-to-CuPy conversion, and more fine-grained control, e.g.,
per-array decision, will be addressed in future work.

To generate Ray-based distributed code from high-level {\em pfor}
loop, the polyhedral phase provides the following data access
information.

\tts{pfor (output = \{$var_{out_1}$ : $type_{out_1}$, $var_{out_2}$ : $type_{out_2}$, $...$\},}

\hspace{10mm}\tts{input = \{$var_{in_1}$ : $type_{in_1}$, $var_{in_2}$ : $type_{in_2}$, $...$\},}

\hspace{10mm}\tts{transfer = $module\_name$)}

\noindent
The \tts{output} and \tts{input} clauses respectively specify the
produced and referenced variables by the {\em pfor} loop and their
corresponding types, while \tts{transfer} clause indicates the
possibility of NumPy-to-CuPy conversion based on the polyhedral
dataflow analysis and library compatibility.
Due to the space limit, the generated code with Ray tasking APIs is omitted.

\subsection{Important Packages Used in Tool Chain Implementations}\label{toolchain}
Our \system compilation flow is built on top of the Python Typed AST
package~\cite{typed-ast}, which serves as the baseline IR to perform
fundamental program analyses and transformations such as type
inference, loop invariant code motions, and constant propagations.
For the polyhedral optimizations presented in
Section~\ref{sec:polyhedral}, we employ \ttt{islpy} package, the Python
interface to the Integer Set Library (ISL)~\cite{verdoolaege2010isl}
for manipulating sets and relations of integer points bounded by
linear constraints.
Beside the polyhedral representations using \ttt{islpy}, we employ
\ttt{sympy}~\cite{Sympy} to manage symolic expressions observed in
the Typed AST.

\section{Experimental Results}\label{sec:experiment}
\subsection{Experimental Setup}

\begin{table}[t]
\caption{Hardware Platform Information (per node) and software versions}
\hspace{-5mm}
\label{tbl:hw-sw-info}
\resizebox{\textwidth}{!}{  
\begin{tabular}{|l|l|l|l|}
\hline
 Per node              & Cori-GPU                           & Summit                          & Titan Xp (workstation)        \\ \hline
CPU                    & 2 $\times$ Intel Xeon Gold 6148    & 2 $\times$ IBM POWER9           & 1 $\times$ Intel i5-7600 CPU  \\
                       & @ 2.40 GHz (40 cores/node)         & @ 3.1 GHz (44 cores/node)       & @ 3.50GHz (4cores)            \\ \hline
GPU                    & 8 $\times$ NVIDIA Tesla V100       & 6 $\times$ NVIDIA Tesla V100    & 1 $\times$ NVIDIA Pascal      \\ \hline
Memory                 & 384GB                              & 512GB                           & 15GB                          \\ \hline
Interconnect           & InfiniBand + PCIe (CPUs-GPUs)      & InfiniBand                      & N/A                           \\
                       &  + NVLink(GPUs)                    & + NVLink (CPUs-GPUs, GPUs)      &                               \\ \hline
Python / NumPy / CuPy  & 3.7.3 / 1.16.4 / 7.4.0             & 3.7.3 / 1.16.0 / 7.4.0          & 3.6.9 / 1.19.5 / 7.2.0        \\ \hline
Ray                    & 0.8.4                              & 0.7.7                           & 0.8.4                         \\ \hline
\end{tabular}
}
\end{table}

We use a standard GPU-equipped workstation, Titan Xp, for single-node
experiments (Section~\ref{sec:result-single}) and two leading HPC
platforms, NERSC Cori~\cite{cori} and OLCF Summit~\cite{summit}
supercomputers, for multi-node experiments
(Section~\ref{sec:result-multi}).  The single-node specification for
each platform is summarized in Table~\ref{tbl:hw-sw-info}.
For Summit, we manually build Ray and its dependencies from scratch
because there is currently no out-of-the-box Python Ray package for
POWER.

\subsection{Single-node Results (Polybench)}\label{sec:result-single}
We first evaluate the impact of our polyhedral optimizations using
PolyBench-Python~\cite{10.1145/3446804.3446842}, which is the Python
implementation of PolyBench~\cite{polybench}, a widely used benchmark
kernels for compiler evaluations.  We use total 15 benchmarks shown
in Table~\ref{tbl:polybench-base-time}, which are appropriate to
evaluate the current library-oriented optimization strategy, while the
evaluation of other 15 benchmarks will be addressed in our on-going
work on hybrid Python/C++ code generation.

\begin{table}[t]
\begin{center}
\caption{PolyBench-Python baselines: Execution time in second (dataset = large)}
\label{tbl:polybench-base-time}
\hspace{-5mm}
\resizebox{\textwidth}{!}{  
\begin{tabular}{|l|c|c|c|c|c|c|c|c|}
\hline
 & 2mm & 3mm & atax & bicg & correlation & covariance & doitgen & gemm \\
\hline
\tabspace{List Default [sec]} & 224.4 & 356.2 & 0.6578 & 0.6730 & 152.5 & 305.7 & 54.46 & 147.4 \\
\hline
\tabspace{List Pluto [sec]} & 205.2 & 337.9 & 0.8381 & 0.8304 & 152.1 & 153.8 & 54.45 & 191.5 \\
\hline
\tabspace{NumPy [sec]} & 0.0214 & 0.03252 & 0.002516 & 0.002447 & 2.212 & 3.813 & 0.1250 & 0.01789 \\
\hline
\hline
 & gemver & gesummv & mvt & symm & syr2k & syrk & trmm & \\
\hline
\tabspace{List Default [sec]} & 1.510 & 0.3068 & 0.8710 & 140.4 & 171.4 & 96.66 & 91.10 & \\
\hline
\tabspace{List Pluto [sec]} & 1.453 & 0.3154 & 0.8714 & 140.5 & 137.9 & 81.73 & 93.27 & \\
\hline
\tabspace{NumPy [sec]} & 0.04676 & 0.001074 & 0.002537 & 1.656 & 2.667 & 0.7839 & 0.8499 & \\
\hline
\end{tabular}
}
\end{center}
\end{table}

\begin{figure}[t]
\begin{center}
\includegraphics[width=\columnwidth]{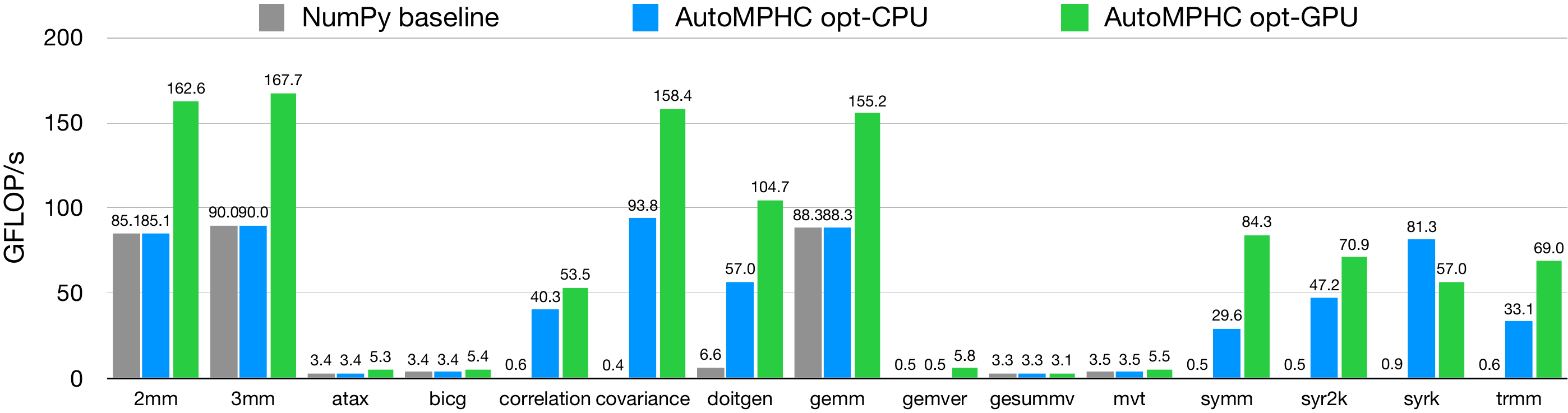}
\end{center}
\vspace{-5mm}
\caption{PolyBench-Python performance on NVIDIA Titan Xp (dataset = extra large)}
\label{fig:polybench-titanxp}
\end{figure}

PolyBench Python provides a variety of benchmark implementations,
including default List version, optimized List version by the Pluto
polyhedral compiler~\cite{Bondhugula:2016:PAP:2914585.2896389}, and
NumPy version.  Table~\ref{tbl:polybench-base-time} shows the
execution time of these versions using ``large'' dataset.  While the
Pluto optimization improves the performance, NumPy version largely
outperforms List versions for all benchmarks.

In the following experiments, we use NumPy version as the baseline of
our comparison, and ``extra large'' dataset to ensure sufficient
execution time.
Figure~\ref{fig:polybench-titanxp} shows the GFLOP/s of three
experimental variants:
\begin{itemize}
\item NumPy baseline: the original NumPy implementation from PolyBench.
\item \system opt-CPU: the CPU optimized version by \system framework.
\item \system opt-GPU: the GPU optimized version by \system framework
  enabling NumPy-to-CuPy conversion.
\end{itemize}

\noindent
Comparing the NumPy and \system opt-CPU versions, our polyhedral
optimization gives 8.7$\times$ -- 212.4$\times$ performance
improvements for \ttt{correlation}, \ttt{covariance}, \ttt{doitgen},
\ttt{symm}, \ttt{syr2k}, \ttt{syrk}, and \ttt{trmm}, while showing
comparable performance for other benchmarks.  Enabling NumPy-to-CuPy
conversion further improves the performance for most benchmarks, with
two exceptions of \ttt{gesummv} and \ttt{syrk}.  In this evaluation,
our profitability conditions always selected GPU variants.  The
improvement of CPU/GPU selection based on offline profiling is an
important future work.

\subsection{Multi-node Results (STAP)}\label{sec:result-multi}

We demonstrate the multi-node performance of our \system compiler
framework using one of our target applications in the signal
processing domain, namely the Space-Time Adaptive Processing (STAP)
application for radar systems~\cite{stap-radar2014}.
The problem size used for STAP is to evaluate the analysis of 144
data cubes for the CPU case; and 2304 data cubes for the GPU case,
where each data cube has \# pulses per cube = 100, \# channels = 1000,
and \# samples per pulse = 30000.  The throughput required for
real-time execution is 33.3 [cubes/sec].
There are three experimental variants as listed below:
\begin{itemize}
\item Python NumPy: The original single-node CPU implementation.
\item Python CuPy: CuPy-based single-node GPU implementation,
  manually ported from the original Python NumPy version.
\item \system: Automatic parallelization by the \system compiler of
  the original Python NumPy version, running on the Ray distributed
  runtime.
\end{itemize}

\begin{figure}[t]
\begin{center}
\includegraphics[width=\columnwidth]{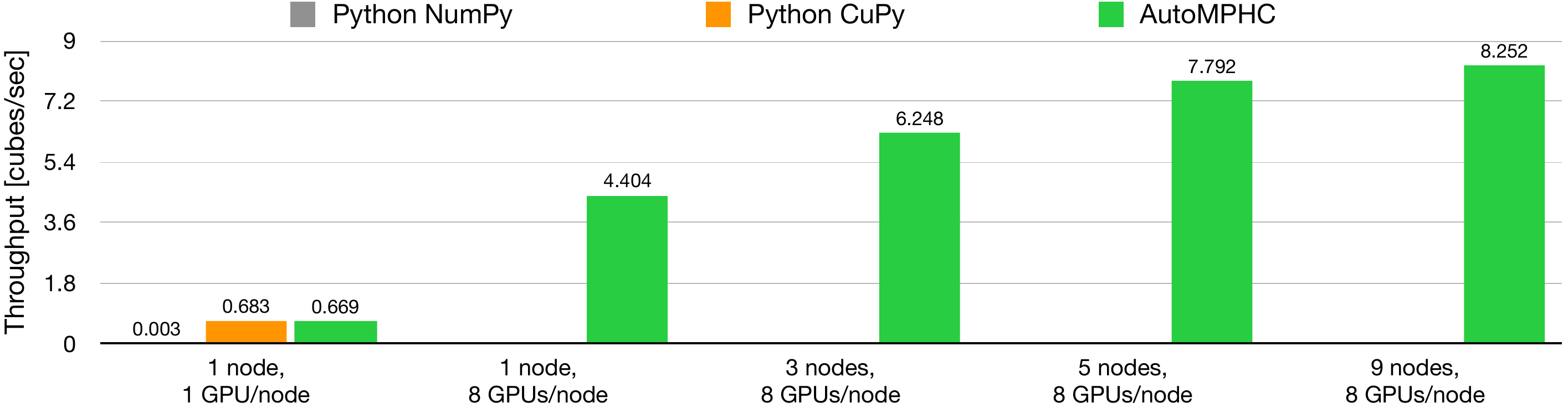}
\end{center}
\vspace{-5mm}
\caption{STAP radar application performance on NERSC Cori supercomputer}
\label{fig:stap-cori}
\begin{center}
\includegraphics[width=\columnwidth]{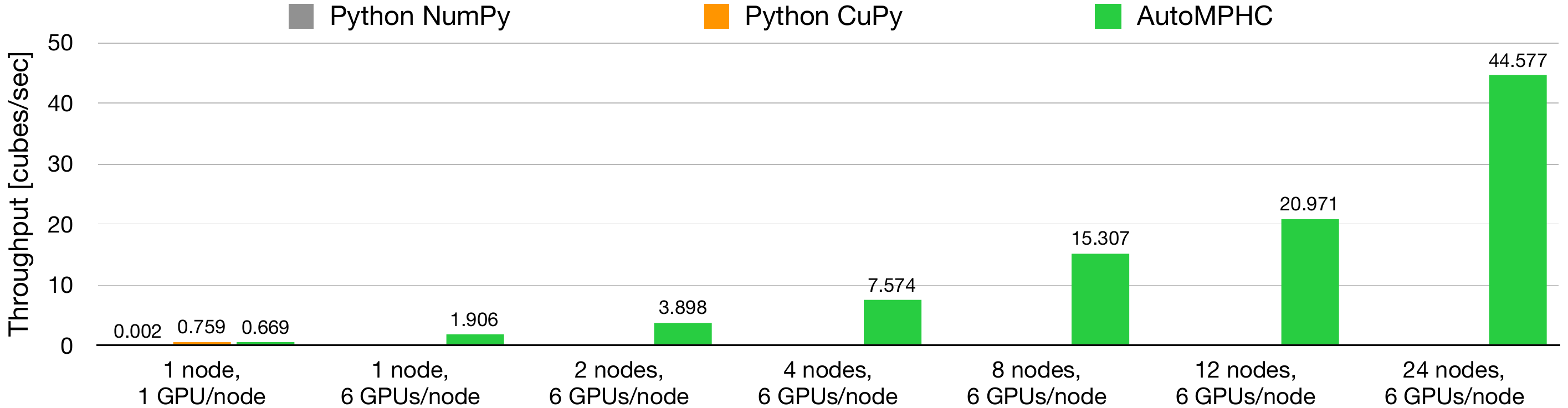}
\end{center}
\vspace{-5mm}
\caption{STAP radar application performance on OLCF Summit supercomputer}
\label{fig:stap-summit}
\end{figure}

\noindent
Figures~\ref{fig:stap-cori} and \ref{fig:stap-summit} show the
throughput performance, i.e., number of data cubes processed per
second, respectively on Cori and Summit.  Given the Python NumPy
version as input, the \system compiler automatically parallelized the
major computation kernel and mapped to GPUs via NumPy-to-CuPy
conversions.  This significantly improves the throughput performance
of the baseline Python NumPy version, while obtaining comparable
performance with the manually ported CuPy implementation when using a
single GPU.
The \system version also shows good multi-node scalability based on
the Ray distributed runtime up to 44.58 [cubes/sec] using 24 nodes on
Summit, which satisfies the domain-specific throughput requirement of
33.3 [cubes/sec] in the real-time scenario of actual radar systems.
The \system version also shows good single-node performance on Cori,
4.40 [cubes/sec], but the multi-node scalability on Cori is more
limited than on Summit.  One reason for this could be the difference
in networks, i.e., Summit's NVLink (50GB/s) vs. Cori's PCIe 3.0 (16GB/s).  In the
parallelized code by \system, each parallel task performs the
computation on the GPU-side and returns the result via the
device-to-host data (D2H) transfers.  We expect that the low-latency
D2H transfer on NVLink contributed to the good scalability of the
\system-generated code on the Summit system.

\section{Conclusions}
This paper describes  \system{} ---a programming system designed
to deliver the benefits of distributed heterogeneous hardware
platforms to domain scientists who naturally use  high-productivity languages like Python.  In our approach, the  parameters and return values of kernel Python functions are annotated with type hints, manually by users or
automatically by profiling tools.
Based on these type hints, the \system compiler performs automatic
AOT parallelization, based on  
advanced polyhedral optimizations, 
CuPy-driven GPU code generation, and 
Ray-targeted heterogeneous distributed code generation and execution.  
The correctness of our AOT parallelization is guaranteed by  multi-version code generation,
since code versions with type-specific optimizations are executed
only when the actual runtime types match  the type hints.
Our empirical evaluations using PolyBench-Python for workstation
performance and the STAP radar application for heterogeneous
distributed performance show significant performance improvements, e.g., up to 358$\times$ improvement for PolyBench
and up to 20,000$\times$ improvement for the STAP radar application,
relative to baseline NumPy-based implementations.
Opportunities for future work include hybrid Python/C++ code generation,
fine-grained NumPy-to-CuPy conversion, and profile-based CPU/GPU
runtime selection.

\bibliographystyle{splncs04}
\bibliography{references.bib}

\begin{thebibliography}{10}
\providecommand{\url}[1]{\texttt{#1}}
\providecommand{\urlprefix}{URL }
\providecommand{\doi}[1]{https://doi.org/#1}

\bibitem{scipy}
Scipy. https://www.scipy.org/ (2001)

\bibitem{numpy}
Numpy. https://numpy.org/ (2006)

\bibitem{cori}
{NERSC Cori Supercomputer}. https://docs.nersc.gov/systems/cori/ (2016)

\bibitem{summit}
Lcf summit supercomputer.
  https://www.olcf.ornl.gov/olcf-resources/compute-systems/summit/ (2019)

\bibitem{typed-ast}
Python typed ast package. https://pypi.org/project/typed-ast/ (2019)

\bibitem{10.1145/3446804.3446842}
Abella-Gonz\'{a}lez, {et al.}: Polybench/python: Benchmarking python
  environments with polyhedral optimizations. In: Proc. of CC 2021 (2021)

\bibitem{julia}
Bezanson, J., Karpinski, S., Shah, V.B., Edelman, A.: Julia: {A} fast dynamic
  language for technical computing. CoRR  \textbf{abs/1209.5145} (2012)

\bibitem{Bondhugula:2016:PAP:2914585.2896389}
Bondhugula, U., Acharya, A., Cohen, A.: The pluto+ algorithm: A practical
  approach for parallelization and locality optimization of affine loop nests.
  ACM Trans. Program. Lang. Syst.  \textbf{38}(3) (Apr 2016)

\bibitem{mpi4py}
Dalcin, L., Fang, Y.L.L.: mpi4py: Status update after 12 years of development.
  Computing in Science Engineering  (2021)

\bibitem{stap-radar2014}
Melvin, W.L.: Chapter 12: Space-time adaptive processing for radar. Academic
  Press Library in Signal Processing: Volume 2 Comm. and Radar Signal Proc.
  (2014)

\bibitem{Ray18}
Moritz, P., Nishihara, R., Wang, S., Tumanov, A., Liaw, R., Liang, E., Elibol,
  M., Yang, Z., Paul, W., Jordan, M.I., Stoica, I.: Ray: A distributed
  framework for emerging ai applications. In: Proc. of OSDI'18 (2018)

\bibitem{polybench}
Poly{B}ench: The polyhedral benchmark suite,
  {\url{http://www.cse.ohio-state.edu/~pouchet/software/polybench/}}

\bibitem{numerical-recipes}
Press, W.H., Teukolsky, S.A., Vetterling, W.T., Flannery, B.P.: Numerical
  Recipes 3rd Edition: The Art of Scientific Computing. 3 edn. (2007)

\bibitem{python}
Python. https://www.python.org/ (1991)

\bibitem{ShHS17}
Shirako, J., Hayashi, A., Sarkar, V.: {Optimized Two-Level Parallelization for
  GPU Accelerators using the Polyhedral Model}. In: Proc. of CC 2017 (2017)

\bibitem{ShPS14}
Shirako, J., Pouchet, L.N., Sarkar, V.: {Oil and Water Can Mix: An Integration
  of Polyhedral and AST-based Transformations}. In: Proc. of SC'14 (2014)

\bibitem{ShSa19}
Shirako, J., Sarkar, V.: {Integrating Data Layout Transformations with the
  Polyhedral Model}. In: Proc. of IMPACT 2019 (2019)

\bibitem{Shirako-lcpc2020}
Shirako, J., Sarkar, V.: An affine scheduling framework for integrating data
  layout and loop transformations. In: Proc. of LCPC 2020 (2020)

\bibitem{Sympy}
Sympy. https://www.sympy.org (2017)

\bibitem{verdoolaege2010isl}
Verdoolaege, S.: {isl: An Integer Set Library for the Polyhedral Model}. In:
  Mathematical Software – ICMS 2010 (2010)

\bibitem{Verdoolaege:2013:PPC:2400682.2400713}
Verdoolaege, S., {et al.}: Polyhedral parallel code generation for {CUDA}. ACM
  Trans. Archit. Code Optim.  \textbf{9}(4),  54:1--54:23 (2013)

\bibitem{lineage-sosp19}
Wang, S., Liagouris, J., Nishihara, R., Moritz, P., Misra, U., Tumanov, A.,
  Stoica, I.: Lineage stash: Fault tolerance off the critical path. In: Proc.
  of the ACM Symposium on Operating System Principles (SOSP'19). p. 338–352.
  SOSP '19 (2019)

\bibitem{Intrepydd-onward20}
Zhou, T., Shirako, J., Jain, A., Srikanth, S., Conte, T.M., Vuduc, R., Sarkar,
  V.: Intrepydd: Performance, productivity and portability for data science
  application kernels. In: Proc. of Onward! '20 (2020)

\bibitem{Zinenko:2018:MCD:3178372.3179507}
Zinenko, O., {et al.}: Modeling the conflicting demands of parallelism and
  temporal/spatial locality in affine scheduling. In: Proc. of CC 2018 (2018)

\end{thebibliography}

\end{document}